\documentclass[aps,prb,twocolumn,showpacs]{revtex4}
\usepackage{graphicx}

\newcommand{\etc}{{\it etc.}}
\newcommand{\etal}{{\it et al.} }
\newcommand{\jc}{$j_c$ }

\newcommand{\SUST}{{\it Supercond. Sci. Technol.}\ }
\newcommand{\JAP}{{\it J. Appl. Phys.}\ }
\newcommand{\APL}{{\it Appl. Phys. Lett.}\ }
\newcommand{\PC}{{\it Physica }C\ }

\newcommand{\EQ}[1]{Equation\,(\ref{#1})}
\newcommand{\FIG}[1]{Figure\,\ref{#1}}

\begin{document}

\title[Magnetic moment of welded HTS samples]
{Magnetic moment of welded HTS samples:\\ dependence on the
current flowing through the welds}

\author{A~B~Surzhenko\footnote[1]{The author to whom correspondence
should be addressed. On leave from Institute for Magnetism, Kiev,
Ukraine}, M~Zeisberger, T~Habisreuther, D~Litzkendorf and
W~Gawalek}
\address{Institut f\"ur Physikalische Hochtechnologie, Winzerlaer Str.
10, Jena D-07745, Germany}

\begin{abstract}
We present a method to calculate the magnetic moments of the
high-temperature super\-con\-ducting (HTS) samples which consist
of a few welded HTS parts. The approach is generalized for the
samples of various geometrical shapes and an arbitrary number of
welds. The obtained relations between the sample moment and the
density of critical current, which flows through the welds, allow
to use the magnetization loops for a quantitative characterization
of the weld quality in a wide range of temperatures and/or
magnetic fields.
\end{abstract}

\pacs{74.72 Bk, 74.80 Bj}


\date{\today}
\maketitle

\section{Introduction}

Practical applications of high-temperature superconductors (HTS)
are based on the HTS ability to generate and to carry a current
which produces magnetic field trying to compensate an external
field \cite{Murakami}. Thus, the HTS performance may be improved
by increasing both the critical current density \jc and the length
scale $d$ over which a current flows \cite{Bean}. An introduction
of the melt-textured (MT) growth process \cite{MTG} generally
allowed to escape large-angle grain boundaries and to reach
thereby quite suitable values of \jc ($\approx 10^5\,A/cm^2$ at
$T=77\,K$). So, over the last decade one spared no efforts to
enlarge the sizes of HTS domains. Since a conventional growth of
extra-large MT crystals \cite{ISTEC} usually requires higher
growth temperature (to avoid parasite-grain nucleation) and,
hence, continues a very long time, various techniques to join
separate HTS blocks were recently proposed \cite{Schatzle,Philip,%
Zheng98,Zheng99,Karapetrov,Freyhardt,Harnois,Puig,Bradley,Noudem,%
Prikhna}. It stimulated a rapid development of experimental
methods describing a quality of the super\-con\-ducting joint,
i.e. the density $j_w$ of critical current which may flow through
it. Direct transport current measurements
\cite{Philip,Bradley,Noudem}, levitation force technique
\cite{Zheng98,Zheng99,Kord}, magneto-optical image analysis
\cite{Zheng99,Puig} and the Hall-probe magnetometry
\cite{Philip,Karapetrov,Freyhardt,Harnois,Puig,Prikhna} were
already used for these purposes.

Each of these methods has its merits and faults. Direct transport
measurements, for example, are suitable only for relatively small
and/or bad welds which critical current $I\sim j_wS$ still does
not heat the sample because of the ohmic losses in the current
pads. The common problems for the other, contactless techniques
are their semi-quantitative nature and low functionality in strong
magnetic fields. The scanning Hall-probe magneto\-metry, for
example, to which these problems regard in the least degree, seems
too sensitive to the test-bench geometry. Since the currents,
which flow on different distances under the scanned sample
surface, ``smeared'' a distribution of magnetic flux density above
the weld, experimental data appear quite hard to interpret
\cite{Matthias} unless, certainly, the sample has the ideal
geometric shape, i.e. flat, thin ring \cite{Zheng99,Karapetrov}.

Meanwhile, both problems do not appear at all for the usual
vibrating-sample magnetometry. The reason why it has so far mainly
used for qualitative description of superconducting joints
\cite{Philip,Zheng99} and not gained a respectable reputation of
an exact, quantitative method may be the following. Re-calculation
of the critical current flowing through the weld(s) requires to
know the magnetic moment of the sample divided onto parts/grains.
In this work we present simple analytical equations for magnetic
moments of split samples having some, most popular geometrical
shapes. The approach is generalized for the case of $n$-slits and
arbitrary sizes.

\section{Results and discussion}

\begin{figure}[!t] \begin{center}
\includegraphics[angle=-90,width=\columnwidth]{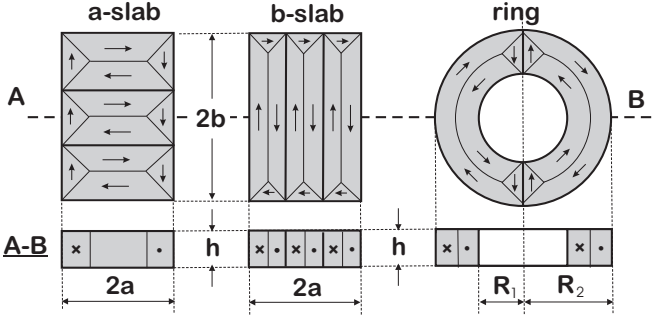}
\end{center} 
\caption{The sketch presents the sample geometries and schematic
distribution of critical currents for the particular case of two
slits, i.e. $n=2,\ f=0$. Depending on which side, either long $2b$
or short $2a$ one, appears parallel to the slits, the slabs are
named $b$-slabs and $a$-slabs, respectively.} \label{Shapes}
\end{figure}

To generalize the estimation procedure for a quality
$f=j_{w}/j_{m}$ of superconducting joints, where $j_w$ and $j_m$
are the densities of the critical current which may flow through a
weld (intra\-grain current) and the original HTS material
(inter\-grain current), let us first introduce the function
$Q(f)=M(f)/M(f=1)$ which numerator and denominator are the
magnetic moments of the welded samples with $j_{w}\neq j_{m}$ and
$j_{w}= j_{m}$, respectively. One can readily predict that $Q$
generally depends on the sample geometry as well as on a quantity
$n$ and a quality $f$ of superconducting joints. Having no aims to
embrace the whole variety of shapes, we shall consider three
common types presented in \FIG{Shapes}, viz., the slabs
split/welded along their short ($2a$) and long ($2b$) sides, which
will hereafter be named $a$- and $b$-slabs, and the ring. By
introducing the asymmetry factor
\begin{eqnarray}
w=\left\{ \begin{array}{ll} a/b, & \textrm{slabs} \\
(R_2-R_1)/(R_2+R_1), & \textrm{ring} \\
\end{array} \right. \label{w}
\end{eqnarray}
\noindent we intend to calculate the functions $Q(w,n,f)$ and to
answer thereby the questions: \begin{description} \item[(Q1)] What
geometry is the more sensitive to the current flowing through the
joint(s)? \item[(Q2)] How to reconstruct its density $j_{w}$
proceeding from magnetic moment of the sample?
\end{description}

To obtain $Q(w,n,f)$ in simple analytical form which could be
convenient for the experimental $j_{w}$ estimation, the following
assumptions will be taken into account:
\begin{description}
\item[(A1)] the Bean's critical state is valid, i.e. the
currents penetrate all over the sample, their density everywhere
equals to the critical current density of the HTS material which
is independent of applied magnetic field $j_{m}(B)=Const$  and,
finally, there is no flux creep.
\item[(A2)] the joints are homogeneous over the whole joining
surface, they have no width and separate the sample onto identical
parts/grains;
\item[(A3)] the welding process does not change the critical current
density $j_{m}$ of the ma\-te\-rial itself.
\end{description}

\subsection{Search for the preferable geometry}
Since the joints were accepted to have no width, no additional
increase of magnetic moment should be expected when $j_{w}$
exceeds $j_{m}$ ($f\geq 1$). The functions $Q(w,n,f)$ must,
therefore, cover the range between their minima, $Q(w,n,f=0)$, and
unity. The more wide this range, the more sensitive appears a
certain geometry to the current flowing through the joint(s). So,
let us estimate the values of $Q(w,n,f=0)$ for chosen geometrical
shapes.

Provided the assumptions (A1) are valid, the magnetic moment of
homogeneous slab ($n=0$ or $f=1$) is presented
\cite{Gyorgy,Murakami} by the simple equation
\begin{equation}
M(1)=V_sj_{m} a\left[1-w/3\right]/2, \label{slab1}
\end{equation}
\noindent where $V_s=2a\times 2b\times h$ is its volume and
$w=a/b\leq 1$ is its asymmetry factor. Since $n$ slits ($j_{w}=0$)
were pre\-sumed to split the original slab onto $N=n+1$ identical
grains, the total moment of the split sample equals $N$ magnetic
moments of each separate part. Thus, for $b$-slabs one can readily
write
\begin{eqnarray}
M(0)&=&V_sj_{m}a\left[1-w/(3N)\right]/(2N),\label{Mb-slab} \\
Q(0)&=&\frac{3N-w}{(3-w)N^2}.\label{Db-slab}
\end{eqnarray}
Taking into account that numerous splitting of the $a$-slab may
reverse long $2b/N$ and short $2a$ sides of its parts, one has to
consider two cases
\begin{eqnarray}
M(0)&=&V_s\frac{j_{m}}{2}\frac{b}{N}%
\left[1-\frac{1}{3wN}\right],\ wN\geq 1 \nonumber \\
M(0)&=&V_s\frac{j_{m}}{2}a%
\left[1-\frac{wN}{3}\right],\  wN\leq 1\label{Ma-slab}
\end{eqnarray}
\noindent which give
\begin{eqnarray}
Q(0)&=&\frac{3wN-1}{(3-w)w^2N^2},\ wN\geq 1 \nonumber \\
Q(0)&=&\frac{3-wN}{3-w},\  wN\leq 1 \label{Da-slab}
\end{eqnarray}

Homogeneous ring ($n=0$ or $f=1$) may naturally be approximated by
a super\-conducting turn which height, inner and outer radii are,
respectively, $h$, $r_i$ and $r_o$. The turn cross-section
$h(r_o-r_i)$ allows to carry a current $I=j_{m} h(r_o-r_i)$. The
turn effective area $\bar{S}=\pi\bar{r}^2$ and its effective
radius $\bar{r}$ are easy to obtain by standard averaging
procedure of the function $S(r)=\pi r^2$ over the range $r_i\leq
r\leq r_o$
\begin{eqnarray}
\bar{S}&=&\left(\int_{r_i}^{r_o}S(r) dr\right)
/(r_i-r_o)=\frac{\pi}{3}\cdot\frac{r_i^3- r_o^3}{r_i-
r_o},\nonumber\\ \bar{r}&=&\sqrt{(r_i^2+r_i r_o+ r_o^2)/3}.
\label{turn}
\end{eqnarray}
\noindent On substituting $r_i=R_1$ and $r_o=R_2$, the ring
magnetic moment $M=I\bar{S}$ fits the equation
\begin{equation}
M(1)=\pi h j_{m} (R_2^3-R_1^3)/3, \label{Mring}
\end{equation}
that appears from the well-known expression \cite{Brandt}
\begin{equation}
M=\pi \int_{-h/2}^{+h/2} dz \int_{R_1}^{R_2} r^2 \cdot j_c(r,z)
dr, \label{axial}
\end{equation}
describing any conductors which remain invariant to their rotation
around the $z$-axis, e.g. toroids, disks, \etc

Similar approach may be applied to split rings or, at least, to
those of them which are thin enough $w\ll 1$ and contain
$n<\pi/(2w)$ slits. \FIG{Shapes} clearly shows that slits separate
such rings onto two nearly full turns which carry the same
circular currents $I=j_{m}h(R_2-R_1)/2$, but flowing in opposite
directions. Since non-circular currents on the left and right
sides of each slit (the diamond-shaped areas in \FIG{Shapes}) are
anti\-parallel and, hence, nearly compensate each others, the
magnetic moment of the split ring may be written as
$M(0)=I(\bar{S_1} - \bar{S_2})[1-n\delta]$, where the small
amendment $\delta\approx w/(2\pi)$ responds for reduction of the
effective turn surfaces around each of $n$-slits. Substituting the
inner and outer radii of the turns which are, respectively, equal
to $R_1$ and $(R_1+R_2)/2$ (for inner turn) and $(R_2+R_1)/2$ and
$R_2$ (for outer one) into Equations (\ref{turn}), one has
\begin{eqnarray}
M(0)&=&\frac{1}{4}V_rj_{m}(R_2-R_1)\left[1-\frac{wn}{2\pi}
\right], \label{Mturns}
\\ Q(0)&=&\frac{3}{4}\cdot
\frac{k^2-1}{k^2+k+1}\left[1-\frac{wn}{2\pi}\right],
\label{Dturns}
\end{eqnarray}
\noindent where the ring volume $V_r$ and its relative thickness
$k$ are, respectively, $\pi(R_2^2-R_1^2)h$ and
$k=R_2/R_1=(1+w)/(1-w)$.

Split rings may also be approximated by replacing the radial arcs
between the slits (see \FIG{Shapes}) with slabs of the same sizes,
viz., $(R_2-R_1)$ and $\pi(R_2+R_1)/n$. Then, by analogy with
Equations (\ref{Ma-slab}) and (\ref{Da-slab}), one writes
\begin{eqnarray}
M(0)&=&\frac{\pi}{4n}V_rj_{m}(R_2+R_1)\left[1-\frac{\pi}{3nw}%
\right],\ nw\geq \pi \nonumber \\
M(0)&=&\frac{1}{4}V_rj_{m}(R_2-R_1)\left[1-\frac{wn}{3\pi}\right],\
nw\leq \pi \label{MRing=Slab} \\
Q(0)&=&\frac{3\pi}{4n}\cdot\frac{(k+1)^2}{k^2+k+1}\left[%
1-\frac{\pi}{3nw}\right],\ nw\geq \pi, \nonumber
\\ Q(0)&=&\frac{3}{4}\cdot\frac{k^2-1}{k^2+k+1}\left[1-\frac{wn}%
{3\pi}\right],\ nw\leq \pi.  \label{DRing=Slab}
\end{eqnarray}
\begin{figure}[!tb] \begin{center}
\includegraphics[width=0.75\columnwidth]{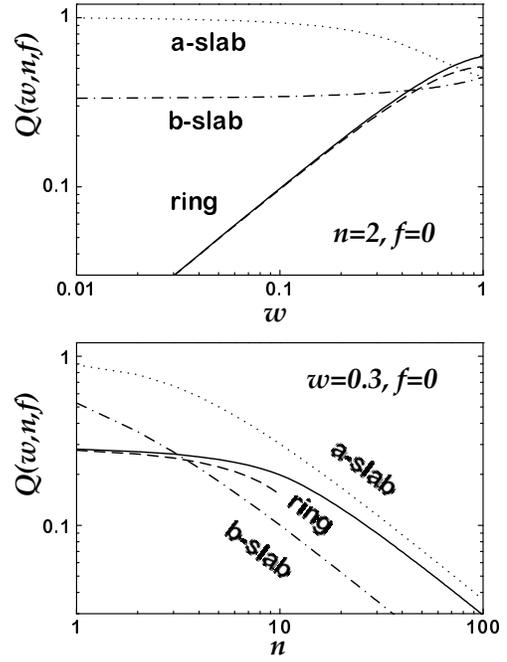}
\end{center}
\caption{Typical dependencies of $Q(w,n,f=0)$ vs the asymmetry
factor $w$ and vs the slit number $n$ confirm that better accuracy
for the density $j_{w}$ of critical current flowing through the
weld(s) will be ensured with the ring ($wn\ll 1$) and the $b$-slab
($wn\gg 1$) geometries of the sample. Double curve for the rings
represent two approximations, viz. the ``opposite turns'' (solid
line) and the ``squared arcs'' (dashed line) approximations given
by the Equations (\ref{Dturns}) and (\ref{DRing=Slab}),
respectively.} \label{fig2}
\end{figure}

Certainly, both the method of ``opposite turns'' (\ref{Dturns})
and that of ``squared arcs'' (\ref{DRing=Slab}) give more accurate
results for relatively thin rings $wn<1$. Fortunately, real rings
usually satisfy this condition. Besides, the $Q(0)$ dependencies
vs the sample asymmetry $w$ and these vs the slit number $n$ (see
\FIG{fig2}) clearly indicate that within the range $wn>1$
definitely lower values of $Q(0)$ and, thus, better accuracy for
the experimental $j_w$ calculation are provided by the $b$-slab
geometry. Thus, further corrections for the rings with $wn\geq 1$
will scarcely have a practical value.

At last, one has to emphasize that the $a$-slabs ensure the worse
accuracy. Moreover, their magnetic moments at $wn<0.1$ are almost
independent of whether the intra\-grain current exists ($f>0$) or
not ($f=0$).

\subsection{Experimental method and its discussion}

We have just obtained analytical solutions for the functions
$Q(w,n,f)$ in the frontier points, $f=0$ and $f=1$. The range
between them, $0<f<1$, may generally \cite{correction} be
represented by the linear combination
\begin{equation}
Q(f)= f Q(1)+(1-f)Q(0), \label{superposition}
\end{equation}
where the coefficient $(1-f)$ appears from an evident condition
that the net (inter\-grain + intra\-grain) current can not exceed
the critical current of the HTS material.

Thus, to estimate the target value of $j_w$, one has to know (i)
magnetic moment of the welded sample, $M(f)$, as well as (ii) the
critical current density $j_m$ of the HTS material. Within the
suggestion (A3), it does not matter which experimental data,
$M(1)$ or $M(0)$, are used to calculate $j_m$. Meantime, heating
of MT HTS crystals up to the temperatures close to their
peritectic points, what actually is the joining procedure
\cite{Schatzle,Philip,Zheng98,Zheng99,Karapetrov,Freyhardt,%
Harnois,Puig,Bradley,Noudem,Prikhna}, may sometimes lead to either
reversible (e.g., the oxygen losses) or irreversible (e.g., the
appearance of cracks and/or their expanding) changes in the HTS
structure. If the latter yet happens, one may, as a last resort,
propose to cut the welded sample again, to measure $M(0)$ and to
recover both the genuine, ``after-the-welding'' value $j_m$ and
the moment $M(1)$, in which this density could result, with aid of
simple relations, (\ref{Mb-slab})-(\ref{Da-slab}) or
(\ref{Mturns})-(\ref{DRing=Slab}). Finally, by substituting the
obtained values, (i) $Q(f)=M(f)/M(1)$ and (ii) $j_m$, into
equation
\begin{equation}
j_w=j_m\frac{Q(f)-Q(0)}{1-Q(0)}, \label{jw}
\end{equation}
we have the density of the current which flows through the
weld(s). Respectively of the obtained quality, either $j_w<j_m$ or
$j_w=j_m$, this current presents the critical current of the weld
or of the HTS itself. The same remains valid for the other
techniques and points out a vital necessity for experimenters to
use a \emph{quantitative} description, rather than to restrict
themselves by qualitative comparison of super\-conducting welds
with their unique HTS material. An ability of our method to
discern the cases $j_w<j_m$ and $j_w=j_m$ and to give, at least,
the lower estimate for the target value $j_w$ seems a worth
compensation for twofold measurements of the magnetic moment.

There are other features, which favorably distinguish this
approach, e.g., its good functionality in wide range of
temperatures and magnetic fields, low requirements for the samples
preparation as well as, generally speaking, the other advantages
which \emph{de facto} turned the magnetic moment measurements into
the world-wide standard to study the super\-conducting properties.

\section*{Acknowledgements}

This work was supported by the German BMBF under the project No
13N6854A3. The authors are grateful to T.~A.~Prikhna for valuable
discussions.



\begin{thebibliography}{50}
\bibitem{correction}
Owing to nonlinear geometry of rings, linear approximation
(\ref{superposition}) is violated. However, a noticeable
correction to the \EQ{superposition} is required only for
relatively thick rings, $0.4\leq w< 1$, at $f\approx 0.5$. For
example, in the ring with $w=1/3$, i.e. where the outer radius
$R_2$ is twice larger than the inner one $R_1$, the nonlinear
amendment does not exceed 0.067
\end{thebibliography}
\end{document}